\documentclass[aps,showpacs,preprint,footinbib,preprintnumbers]{revtex4}
\usepackage{amssymb}
\usepackage{amsmath}
\usepackage{amsfonts}
\usepackage{bm}
\usepackage[dvipdfmx]{graphicx}

\newcommand{\be}{\begin{equation}}
\newcommand{\ee}{\end{equation}}
\newcommand{\bea}{\begin{eqnarray}}
\newcommand{\eea}{\end{eqnarray}}
\newcommand{\del}{\partial}

\newcommand{\nonum}{\nonumber}

\begin{document}

\title{Kaon-Nucleon systems and their interactions in the Skyrme model}

\author{Takashi Ezoe$^1$ and Atsushi Hosaka$^{1,2}$}
\affiliation{$^{1}$Research Center for Nuclear Physics, Osaka University, Ibaraki, 567-0048, Japan}
\affiliation{$^{2}$Advanced Science Research Center, Japan Atomic Energy Agency, Tokai, Ibaraki, 319-1195 Japan}

\date{\today}
\begin{abstract}
We study kaon-nucleon systems in the Skyrme model in a method based on the bound state approach of Callan-Klebanov but with the kaon around the physical nucleon of the rotating hedgehog. This corresponds to the variation after projection, reversing the order of semiclassical quantization of $1/N_c$ expansion. 
The method, however, is considered to be suited to the study of weakly interacting kaon-nucleon systems including loosely $\bar{K}N$ bound states such as $\Lambda(1405)$. 
We have found a bound state with binding energy of order ten MeV, consistent with the observed state.
We also discuss the $\bar{K}N$ interaction and find that it consists of an attraction in the middle range and a repulsion in the short range.

\end{abstract}
\maketitle
\section{Introduction}
Recently, kaon and nucleon systems have been receiving a lot of attention in hadron and nuclear physics.
In particular, the anti-kaon and nucleon $(\bar{K}N)$ interaction is expected to be strongly attractive and 
it is considered that they form a bound state which eventually becomes a resonance by the coupling to the open channel of $\pi \Sigma$.  
The resulting Feshbach  resonance state is identified with $\Lambda \left( 1405 \right)$~\cite{KbarN bound state 1, KbarN bound state 2}, which is the state that cannot be easily explained by a three quark state.  
Based on the basic features of the $\bar{K}N$ properties, there have been large number of discussions in a few body nuclear systems with the kaon as deeply bound states~\cite{few-body system 1, few-body system 2, few-body system 3, few-body system 4, few-body system 5}.
Detailed properties of these few-body systems, however, are yet under debate.  
A possible reason for that is that the kaon-nucleon interaction is not well understood.

Several kaon-nucleon interaction have been derived by a phenomenological method and by chiral theories~\cite{yamazaki 1, yamazaki 2, chiral potential 1, chiral potential 2, chiral potential 3}.
Akaishi and Yamazaki proposed a $\bar KN$ potential with a strong attraction~\cite{yamazaki 1, yamazaki 2}.
Their potential is phenomenological with several model parameters.
The chiral approach is based on the low energy theorem of spontaneously broken chiral symmetry, 
that is the Weinberg-Tomozawa interaction~\cite{WT 1, WT 2}.  
It gives the correct $T$-matrix at low energies, but for resonances it needs unitarization which 
necessarily requires a parameter to regularize the divergence associated with 
the point-like nature of the interaction in the three-dimensional space.  
Furthermore, in the latter approach the concept of potential is not required as long as observed quantities 
are calculated from the $T$-matrix.  
In the calculations of few-body systems, however, the interaction in the form of potential is more convenient.  
%
%

In this article, we derive the kaon-nucleon interaction in the bound state approach of the Skyrme model.
Due to the extended structure of the nucleon as a soliton, the resulting 
interaction can be expressed as a potential.
In the Skyrme model, the nucleon emerges as a soliton of a non-linear field theory of  the pion and then describes the extended structure of the nucleon~\cite{Skyrme model 1, Skyrme model 2, Skyrme model 3}. 
The model contains parameters which are, however,  determined by 
the properties of the nucleon itself or inputs other than the kaon dynamics.  
In this sense our approach is free from parameters.  

Our bound state approach is based on the one proposed by Callan and Klebanov~\cite{Callan-Klebanov 1, Callan-Klebanov 2}, where kaons are introduced as fluctuations around the Skyrmion.  
Their original method followed precisely the $1/N_c$ counting for the quantization of the kaon fluctuations and Skyrmion rotations.  
Kaons are moving around the hedgehog soliton with a fixed orientation.  
Due to the strong attraction of the Wess-Zumino term~\cite{WZ term 1, WZ term 2, WZ term 3}, bound states are generated for the $\bar{K}$-hedgehog systems.  
Because of the coupling of the spin and isospin of the hedgehog, after the quantization, the bound $\bar{K}$ carries spin rather than isospin as the original one does.
Thus the bound $\bar{K}$ is regarded as the strange quark.  
This method provides an interesting picture of the $\bar{K}N$ bound system but it is not suited to the description of the physical kaon and the nucleon.  

In the present paper, we propose an alternative method; we first quantize the hedgehog Skyrmion to generate the physical nucleon and then introduce the physical kaon around it.  
Our method, however, does not obey strictly the $1/N_c$ counting rule, because the hedgehog rotation is of higher order than the kaon fluctuation. 
For the physical situation, however, we consider it reasonable, 
as long as we discuss weakly bound states with binging energy of order ten MeV, which is a typical energy of hadronic scale.  
In such a situation, motion of the kaon is expected to be slower than the motion of the hedgehog rotation. 
Our method is justified if this condition is well satisfied, and corresponds to the variation after projection in many-body physics~\cite{Ring-Schuck}.   

We organize the paper as follows.  
In section 2, we explain our method with some review on the original bound state approach of Callan-Klebanov.  
The difference between their and our methods is explained in detail.
In section 3, we present results of our method for the $\bar{K}N$ bound states.
Then we analyze the $\bar{K}N$ potential.  
Several properties of the resulting potential are investigated.  
In section 4, we summarize the present work and discuss some further studies. 

\section{Method}
\subsection{Skyrme Lagrangian and the new ansatz}
Let us start with the Skyrme Lagrangian~\cite{Skyrme model 1, Skyrme model 2, Skyrme model 3}
\bea
	L = \displaystyle{\frac{1}{16}} F_{\pi}^2 \mathrm{tr} \left( \del_{\mu}U \del^{\mu}U^{\dag} \right)
		+ \displaystyle{\frac{1}{32e^2}} \mathrm{tr}\left[ \left( \del_{\mu}U\right)U^{\dag}, \left( \del_{\nu}U\right) U^{\dag} \right]^2
		+ L_{WZ}
		+ L_{SB},
\label{eq:lagrangian}
\eea
where $U$ is the SU(3)-valued chiral field
\bea
	U = \exp \left[ i \displaystyle{\frac{2}{F_{\pi}}} \lambda_a \phi_a \right] , \ \ \ \ a = 1, 2, 3, \cdots ,8,
\eea
\bea
	\phi = \displaystyle{\frac{1}{\sqrt{2}}} \sum_{a = 1}^8 \lambda_a \phi_a =
		\begin{pmatrix}
			\displaystyle{\frac{1}{\sqrt{2}}} \pi^0 + \displaystyle{\frac{1}{\sqrt{6}}} \eta & \pi^+ & K^+ \\
			\pi^- &  - \displaystyle{\frac{1}{\sqrt{2}}} \pi^0 + \displaystyle{\frac{1}{\sqrt{6}}} \eta & K^0 \\
			K^- & \bar{K}^0 & -\displaystyle{\frac{2}{\sqrt{6}}} \eta
		\end{pmatrix},
\eea
and $\lambda_a$ are the Gell-Mann matrices.
The Wess-Zumino term $L_{WZ}$ is given by~\cite{WZ term 1, WZ term 2, WZ term 3}
\bea
	L_{WZ} = - \displaystyle{\frac{i N_c}{240 \pi^2}} \int d^5 x \  \varepsilon^{\mu \nu \alpha \beta \gamma}
				\mathrm{tr} \left[ \left( U^{\dag} \del_{\mu} U \right) \left( U^{\dag} \del_{\nu} U \right) \left( U^{\dag} \del_{\alpha} U \right) \left( U^{\dag} \del_{\beta} U \right) \left( U^{\dag} \del_{\gamma} U \right) \right],
\eea
where $N_c$ is the number of colors.
In this paper, we set $N_c = 3$. 
The last term in Eq.~(\ref{eq:lagrangian}) $L_{SB}$ is the explicit symmetry breaking term due to the finite masses of pseudo-scalar mesons~\cite{SU(3) Skyrme model 1, SU(3) Skyrme model 2}
\bea
	L_{SB} = \displaystyle{\frac{1}{48}} F_{\pi}^2 \left( m_{\pi}^2 + 2 m_K^2 \right) \mathrm{tr} \left( U + U^{\dag} -2 \right)
			+ \displaystyle{\frac{\sqrt{3}}{24}} \left( m_{\pi}^2 - m_K^2 \right) \mathrm{tr} \left[ \lambda_8 \left( U + U^{\dag} \right) \right].
\eea
In the present paper, we consider the chiral limit for the $u, d$ sector, $m_u = m_d = 0, m_s \neq 0$. 
This means to set $m_{\pi} = 0, m_K \neq 0$.
There are three model parameters, the pion decay constant $F_{\pi}$, the Skyrme parameter $e$, and the mass of the kaon $m_K$.
Their actual values will be discussed in Section~\ref{sec:Results and discussions}.  

Callan and Klebanov introduced the following ansatz (CK ansatz)~\cite{Callan-Klebanov 1, Callan-Klebanov 2}
\bea
	U_{CK} = \sqrt{N} U_K \sqrt{N},
\label{ans:CK ansatz}
\eea
where
\bea
	N 
		= \begin{pmatrix}
			\xi^2 & 0 \\
			0      & 1
		\end{pmatrix}
	, \ \ \ \ \xi^2 = U_{\pi} =  \exp \left[ \displaystyle{\frac{2 i}{F_{\pi}}} \bm{\tau} \cdot \bm{\pi}\right],
\label{eq:def of pionic soliton}
\eea
\bea
	U_K = \exp \left[ \displaystyle{\frac{2 \sqrt{2} i}{F_{\pi}}}
			\begin{pmatrix}
				0             & K \\
				K^{\dag} & 0
			\end{pmatrix}
				\right]
	, \ \ \ \ K = \begin{pmatrix}
				K^{+} \\
				K^{0}
			\end{pmatrix}.
\label{eq:def of kaon fluctuation}
\eea
They followed the $1/N_c$ expansion scheme when constructing the kaon-nucleon system; the hedgehog nucleon is formed in the leading order of $N_c^1$, kaon fluctuations are introduced in the next to leading order of $N_c^0$, and finally the hedgehog-kaon system is rotated in spin-isospin space.
This is to rewrite the ansatz Eq.~(\ref{ans:CK ansatz}) as
\bea
	U_{CK} \rightarrow A \left( t \right) \sqrt{N_H} U_K \sqrt{N_H} A^{\dag} \left( t \right),
\label{eq:CK quantization}
\eea
where $N_H$ denotes the hedgehog configuration
\bea
	N_H =
		\begin{pmatrix}
			\xi^2 & 0 \\
			0       & 1
		\end{pmatrix}
	, \ \ \ \ \xi^2 = U_H =  \exp \left[ i F \left( r \right) \bm{\tau} \cdot \hat{r} \right]
\label{eq:def of Hg configuration} 
\eea
with $F \left( r \right)$ being the soliton profile function, and $A \left( t \right)$ is a time-dependent SU(2) rotation matrix.

By quantizing the rotating system with kaon fluctuations, they have generated the physical hyperons such as $\Lambda$, $\Sigma$ baryons~\cite{Callan-Klebanov 1, Callan-Klebanov 2}.
A unique feature of their method is that there occurs a transmutation between the spin and isospin quantum numbers of the kaon due to the background field of the hedgehog configuration; 
the anti-kaon $( s \bar{u} )$ behaves as a strange quark and the kaon $( \bar{s} u )$ behaves as an anti-strange quark.
One of the purposes of the present paper is to study the interaction between the kaon and the nucleon.
Due to the feature as explained above, the CK ansatz is not convenient for this purpose.
To do that, here, we would like to propose an alternatively ansatz.
First we construct the physical nucleon and then introduce the kaon fluctuations.
This amounts to writing the ansatz
\bea
	U = A \left( t \right) \sqrt{N_H} A^{\dag} \left( t \right) U_K  A \left( t \right) \sqrt{N_H} A^{\dag} \left( t \right),
\label{ans:our ansatz}
\eea
where $N_H$ is and $U_K$ is given by Eq.~(\ref{eq:def of Hg configuration}) and Eq.~(\ref{eq:def of kaon fluctuation}), respectively, and $A \left( t \right)$ is a time-dependent SU(2) matrix as we mention above.

We comment the differences of our ansatz from the CK one. 
In the CK ansatz, the kaon is the fluctuation around the hedgehog soliton. 
Their quantization method Eq.~(\ref{eq:CK quantization}) is based on the picture that the kaon is strongly bound to it.
Contrary,  in our ansatz, the kaon is introduced as the fluctuation around the physical nucleon. 
Thus the hedgehog soliton is first rotated in our ansatz Eq.~(\ref{ans:our ansatz}).
This is based on the picture that the kaon is weakly bound to the nucleon as expected to hadronic molecules.  
This corresponds to the variation after projection in the many-body physics~\cite{Ring-Schuck}.  

\subsection{Kaon fluctuations around the physical nucleon}
\label{sec:equation of motion}
To describe the kaon fluctuations around the physical nucleon, let us first decompose the kaon field as
\bea
	\begin{pmatrix}
		K^+ \\
		K^0
	\end{pmatrix}
	       &=& \psi_I K \left( t, \bm{r} \right) \nonum \\
	       &\rightarrow& \psi_I K \left( \bm{r} \right) \exp \left( - i E t \right),
\eea
where $\psi_I$ is the two component isospinor,
and the spatial wave function $K \left( \bm{r} \right)$ is expanded by the spherical harmonics $Y_{lm} \left( \hat{r} \right)$ 
\bea
	 K \left( \bm{r} \right) = \sum_{\alpha l m} C_{l m \alpha} Y_{lm} \left( \hat{r} \right) k_l^{\alpha} \left( r \right)
\eea
with the expansion coefficients $C_{l m \alpha}$ and the radial wave function $k_l^{\alpha} \left( r \right)$.

Substituting Eq.~(\ref{ans:our ansatz}) for the lagrangian Eq.~(\ref{eq:lagrangian}), we take into account the terms up to second order of the kaon fields.
Taking a variation with respect to the kaon fields, we obtain the equation of motion for the kaon radial wave function $k_l^{\alpha} \left( r \right)$
\bea
	- \displaystyle{\frac{1}{r^2}} \displaystyle{\frac{d}{dr}} \left( r^2 h\left( r \right) \displaystyle{\frac{dk_l^{\alpha} \left( r \right)}{dr}}\right)
		- E^2 f \left( r \right) k_l^{\alpha} \left( r \right)
		+ \left( m_K^2 + V \left( r \right) \right) k_l^{\alpha} \left( r \right) = 0,
\label{eq:equation of motion for kaon}
\eea
where
\bea
	h(r) = 1 + \displaystyle{\frac{1}{\left( e F_{\pi} \right)^2}} \displaystyle{\frac{2}{r^2}} \sin^2 F, 
\label{eq:h}
\eea
\bea
	f(r) = 1 + \displaystyle{\frac{1}{\left( e F_{\pi} \right)^2}} \left( \displaystyle{\frac{2}{r^2}} \sin^2 F + F'^2 \right),
\label{eq:f}
\eea
\bea
	V \left( r \right) = V_0^c \left( r \right) + V_{\tau}^c \left( r \right) I_{KN} + V_0^{LS} \left( r \right) J_{KN} + V_{\tau}^{LS} \left( r \right) J_{KN} I_{KN},
\label{eq:potential}
\eea
and
\bea
	I_{KN} = \bm{I}^K \cdot \bm{I}^N, \ \ \ \
	J_{KN} = \bm{L}^K \cdot \bm{J}^N.
\label{eq:spin and isospin products}
\eea
In Eq.~(\ref{eq:spin and isospin products}), the nucleon spin and isospin operators, $\bm{J}^N$ and $\bm{I}^N$, are given by~\cite{zahed}
\bea
	\bm{J}^N &=& i \Lambda \mathrm{tr} \left[ \bm{\tau} \dot{A}^{\dag} \left( t \right) A \left( t \right) \right], \\
	\bm{I}^N &=& i \Lambda \mathrm{tr} \left[ \bm{\tau} \dot{A} \left( t \right) A^{\dag} \left( t \right) \right],
\eea
where $\dot{A} \left( t \right)$ is the time derivative of $A \left( t \right)$, $\bm{\tau}$ is the $2 \times 2$ Pauli matrices, and $\Lambda$ is the soliton moment of inertia which is given by~\cite{ANW}
\bea
	\Lambda = \displaystyle{\frac{2 \pi}{3}} F_{\pi}^2 \int dr \ r^2 \sin^2 F \left[ 1 + \displaystyle{\frac{4}{\left( e F_{\pi} \right)^2}} \left( F'^2 + \displaystyle{\frac{\sin^2 F}{r^2}} \right) \right].
\eea
The kaon isospin operator, $\bm{I}^K$, is given by the $2 \times 2$ Pauli matrices
\bea
	\bm{I}^K = \displaystyle{\frac{\bm{\tau}}{2}}.
\eea
Lastly, $\bm{L}^K$ in Eq.~(\ref{eq:spin and isospin products}) is the orbital angular momentum operator for the kaon
\bea
	\bm{L}^K = \bm{r} \times \bm{p}^K.
\eea

Using the present ansatz Eq.(\ref{ans:our ansatz}), the resulting lagrangian and equation of motion Eq.~(\ref{eq:equation of motion for kaon}) contain the rotation matrix $A \left( t \right)$ in several places.
In other words, in these equations, terms of different order of $1/N_c$ exist simultaneously, indicating the violation of $1/N_c$ expansion.
This, however, is the feature of the present approach which we consider suited to the study of the physical kaon and nucleon interaction.

We note that the potential Eq.~(\ref{eq:potential}) has four components;
the isospin independent and dependent central forces, $V_0^c$ and $V_{\tau}^c$, respectively, and similarly for the spin-orbit forces $V_0^{LS}$ and $V_{\tau}^{LS}$.
In fact, these terms complete the general structure of the potential between the isospinor-pseudoscalar kaon and isospinor-spinor nucleon.
In Appendix, we write down the explicit expressions of $V \left( r \right)$.

\section{Results and discussions}
\label{sec:Results and discussions}
In this section, we consider kaon and nucleon bound states and their potential.
In our approach, there are three parameters: the pion decay constant $F_{\pi}$, the Skyrme parameter $e$, and the mass of the kaon $m_K$.
We keep $m_K = 495 \  \mathrm{MeV}$, and we consider three parameter sets for $F_{\pi}$ and $e$.
The parameter set 1 is $\left( F_{\pi}, e \right) = \left( 129 \  \mathrm{MeV}, 5.45 \right)$, which is adjusted to fit the masses of the nucleon and the delta~\cite{ANW}.
The parameter set 2 is $\left( F_{\pi}, e \right) = \left( 186 \ \mathrm{MeV}, 5.45 \right)$, where the pion decay constant $F_{\pi}$ is fixed at the experimental value while $e$ is unchanged from the set 1.
The last parameter set 3 is $\left( F_{\pi}, e \right) = \left( 186 \  \mathrm{MeV}, 4.82 \right)$, which is adjusted to fit the mass difference of nucleon and delta with $F_{\pi} = 186 \  \mathrm{MeV}$.

\subsection{Bound states}
\label{sec:bound state}
As discussed by Callan and Klebanov~\cite{Callan-Klebanov 1, Callan-Klebanov 2}, the bound state properties differ for the kaon $( K )$ and the anti-kaon $( \bar{K} )$.
The difference is due to the Wess-Zumino term which provides an attractive interaction for the $\bar{K}$ while repulsive one for the $K$, allowing bound states only for the antikaon-nucleon $( \bar{K} N )$ systems.
This feature still holds in our present approach.
In the following, we consider only the $\bar{K} N$ systems.

To investigate $\bar{K} N$ states, we have solved numerically the equation of motion Eq.~(\ref{eq:equation of motion for kaon}) for various partial waves and isospin.
The kaon and nucleon systems take isospin 0 and 1, and each of them allows total spin and parity $J^P = 1/2^{\pm}, 3/2^{\pm}, \cdots$.
We have studied several low-lying states, and found that bound states exist for $J^P = 1/2^- \left( l  =  0 \right)$.
In fact, this is the lowest bound state as we naively expect, in contrast with the result of Callan and Klebanov.   
The numerical results are summarized in Table~\ref{tab:bound state properties}.

For the parameter set 1, we found one bound state both for $I=0$ and 1, with the binding energies (B.E.) 82.9 MeV and 43.1 MeV, respectively.
The former may be identified with the $\Lambda(1405)$, whose binding energy is, however, too strong.
This is due to the use of the small pion decay constant as compared to the experimental value.
As we will discuss later, the important contribution to the interaction is proportional to $1/F_\pi^2$.
The second bound state may be identified with a $\Sigma$ hyperon.
Experimentally, there are several low lying $\Sigma$ resonances but with only weak significance~\cite{PDG}.
By considering the mass difference of the two bound states, one candidate would be
$\Sigma(1480)$.

For the parameter set 2, we have found one bound state only for $I=0$ with the binding energy 27.2 MeV.
This is significantly weaker than the result of set 1 and leads to the total mass closer to the experimentally observed one of $\Lambda(1405)$.
As mentioned above, the difference is due to the change in the pion decay constant.
It seems that the use of the experimental value of $F_\pi$ is important to reproduce numerically the properties of the kaon and nucleon systems.
For the parameter set 3, the binding energy is 32.9 MeV which is slightly larger, but in the similar order of magnitude to the result of set 2.

To understand better the bound state properties, in Table~\ref{tab:bound state properties}, we show root mean square radii $\left< r^2 \right>^{1/2}$ for the baryon number distribution of the nucleon and for the kaon radial function.
They are defined by
\bea
	\left< r_N^2 \right> &=& \int_0^{\infty} dr \ r^2 \rho_B \left( r \right), \\
	\left< r_K^2 \right> &=& \int_0^{\infty} dr \ r^4 k^2 \left( r \right), 
\eea
where $\rho_B \left( r \right)$ is the baryon charge density and is given by~\cite{ANW} 
\bea
	\rho_B \left( r \right) = - \displaystyle{\frac{2}{\pi}} \sin^2 F F'.
\eea
The baryon number radii are about 0.5 fm corresponding to the nucleon core size, while the kaon wave function extends up to 1 fm, indicating that the kaon is moving around the nucleon with weak binding.
To see a bit more detail of Table~\ref{tab:bound state properties}, we observe that as the binding energy increases, in the order of set 2, set 3, and set 1, the hedgehog (baryon number) distribution increases, while the kaon distribution decreases.
The fact that the bound state extends less for a larger binding energy is consistent with the general property of bound states.

\begin{figure}[htb]
	\begin{center}
		\includegraphics[width = 100mm]{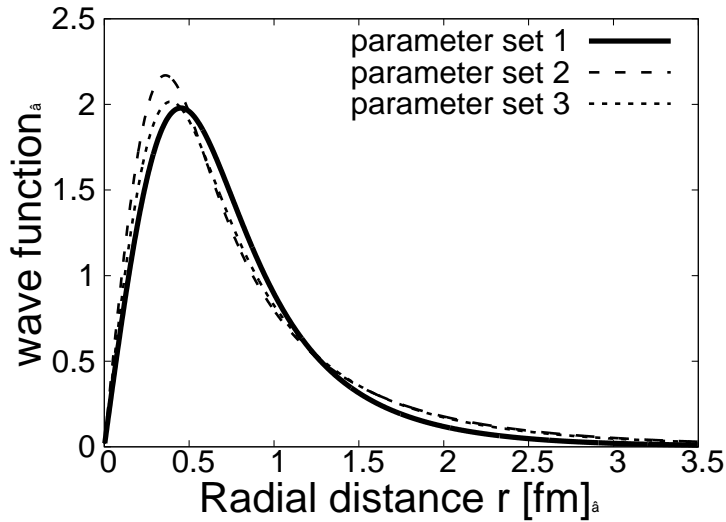}
	\end{center}
	\caption{The wave functions of $\bar{K}N$ bound states $\left( I = 0 \right)$ for three parameter sets in units of $1/\left( e F_{\pi} \right)^{3/2}.$}
\label{fig:wave functions}
\end{figure}

In Figure~\ref{fig:wave functions}, we have shown the normalized kaon wave functions $|k(r)|$ for the three parameter sets.
It is interesting to see that the wave function vanishes at the origin, although it is the s-wave. 
This is due to the presence of the repulsive core in the potential as we will see in the next subsection.
In the large $r$ region, wave functions extends further for smaller binding energies, which explains the behavior of $\left< r^2_K \right>^{1/2}$ depending on the binding energy.
The peak position of the wave function, however, is correlated with the attractive minima of the potential (as shown in the next subsection).
Finally, we would like to emphasize that bound states exist with a binding energy of order ten MeV which is the typical order of hadronic interaction.
This contrasts the Callan-Klebanov's result~\cite{Callan-Klebanov 1, Callan-Klebanov 2, Itzhaki}, as we will discuss in the subsection~\ref{sec:Comparisons with the CK} in detail.
\begin{table}[htb]
	\begin{center}
		\caption{The properties of the $\bar{K}N\left( I = 0 \right)$ bound states}
			\begin{tabular}{| c || c | c | c | c | c |} \hline
	  					          & $F_{\pi}$ [MeV] &  $e$  & B.E. [MeV] & $\left< r^2_{N} \right>^{1/2}$ [fm] & $\left< r^2_{K} \right>^{1/2}$ [fm] \\ \hline
				parameter set 1 & 129                     & 5.45 & 82.9 	       & 0.59 					     & 0.99						  \\ \hline 
				parameter set 2 & 186                     & 5.45 & 27.2 	       & 0.41 				             & 1.19						  \\ \hline
				parameter set 3 & 186        	      & 4.82 & 32.9 	       & 0.46					     & 1.18					  	  \\ \hline
			\end{tabular}
			\label{tab:bound state properties}
	\end{center}
\end{table}

\subsection{Potential}
\label{sec:potential}
In this subsection, we study the potential for the kaon nucleon system.
It has been already defined in the Klein-Gordon like equation~(\ref{eq:equation of motion for kaon}) by $V \left( r \right)$ in Eq.~(\ref{eq:potential}).
Thus the potential $V \left( r \right)$ carries the dimension of MeV$^2$.
Now it is convenient to define an alternative one in units of MeV which is used in a schr\"{o}dinger-like equation.
To do that, we first rewrite Eq.~(\ref{eq:equation of motion for kaon}) in the following form
\bea
	- \displaystyle{\frac{1}{m_K + E}} \displaystyle{\frac{1}{r^2}} \frac{d}{dr} \left( r^2 \frac{dk_l^{\alpha} \left(r\right)}{dr} \right)
		+ U \left( r \right) k_l^{\alpha} \left(r\right)
		= \varepsilon k_l^{\alpha} \left(r\right),
\label{eq:schrodinger equation}
\eea
where
\bea
	E = m_K + \varepsilon,
\eea
and
\bea
	U \left( r \right)
		&=& - \displaystyle{\frac{1}{m_K + E}} \left[ \displaystyle{\frac{h \left( r \right)-1}{r^2}} \displaystyle{\frac{d}{dr}} \left( r^2 \displaystyle{\frac{d}{dr}} \right)
			+ \displaystyle{\frac{d h(r)}{dr}} \displaystyle{\frac{d}{dr}} \right]
			- \displaystyle{\frac{\left( f \left( r \right) - 1 \right) E^2}{m_K + E}}
			+ \displaystyle{\frac{V (r)}{m_K + E}}. \nonum \\
\label{eq:effective potential}
\eea
In Eq.~(\ref{eq:effective potential}), $h \left( r \right)$, $f \left( r \right)$, and $V (r)$ are given by Eq.~(\ref{eq:h}), Eq.~(\ref{eq:f}), and Eq.~(\ref{eq:potential}), respectively.
This potential $U \left( r \right)$ has the following properties.
First, it is nonlocal and depends on the kaon energy.
Second, it contains four components of the isospin independent and dependent, central and LS terms as we mentioned in subsection~\ref{sec:equation of motion}.
Third, near the origin, this potential behaves as a repulsive or an attractive force proportional to $1/r^2$ depending on the total isospin and total spin.
Details of this behavior are discussed in Appendix. 

Because the potential $U \left( r \right)$ formally contains derivative, we make the following equivalent quantity
\bea
	\tilde{U} \left( r \right) \equiv \displaystyle{\frac{U \left( r \right) k_l^{\alpha} \left(r\right)}{k_l^{\alpha} \left(r\right)}}.
\label{eq:potential value}
\eea
In this paper, we computed it by using the bound state wave function.
Therefore, strictly speaking the potential derived here is for $l=0$ bound state.
In principle it is also possible to calculate $\tilde{U} \left( r \right)$ for other $l$'s by using scattering wave functions.
The study of the scattering states will be discussed elsewhere.

\begin{figure}[htbp]
	\begin{center}
		\includegraphics[width = 100mm]{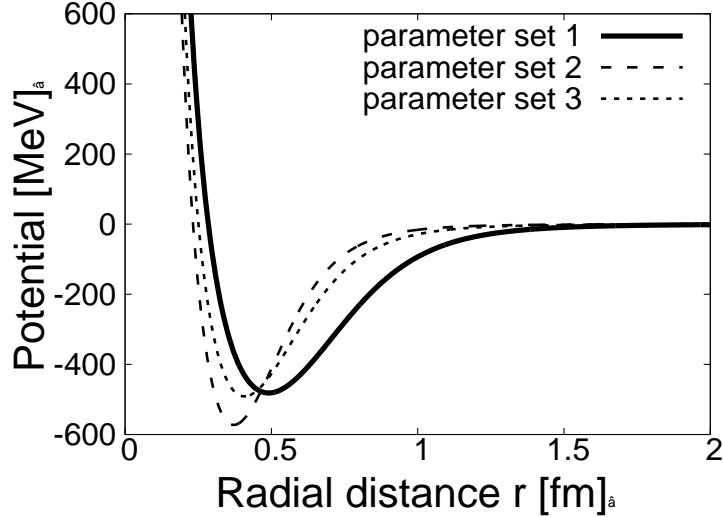}
	\end{center}
\caption{The equivalent potentials $\tilde{U}$ defined in Eq.~(\ref{eq:potential value}) for the $\bar{K} N \left( I = 0 \right)$ bound states.}
\label{fig:potential}
\end{figure}

The resulting  $\tilde{U} \left( r \right)$ is plotted in Figure~\ref{fig:potential} for the three parameter sets as used in the previous subsection.
In the ordering of 1, 3, and 2, the potential minimum moves from outside to inside, and with the potential depth increasing.
In accordance with this change, the shapes of the kaon wave functions have been explained in the previous subsection.

Now,  it is instructive to investigate further properties in comparison with what we expect in the chiral theory.
In the chiral theory, $\bar{K}N$ interaction is derived from the Weinberg-Tomozawa interaction (WT interaction) and is given by the following lagrangian~\cite{WT 1, WT 2}
\bea
	L_{WT} = \displaystyle{\frac{2}{F_{\pi}^2}} \left\{ \bar{N} \bm{I}^N \gamma^{\mu} N \cdot \left( \partial_{\mu} K^{\dag} \bm{I}^K K - K^{\dag} \bm{I}^K \partial_{\mu} K \right) \right\},
\label{eq:WT lagrangian}
\eea
where $\bm{I}^N$ and $\bm{I}^K$ are the nucleon and the kaon isospin operators, respectively.

A feature of Eq.~(\ref{eq:WT lagrangian}) is that the interaction strength is proportional to $1/F_{\pi}^2$.
Therefore, the interaction becomes stronger for smaller $F_{\pi}$ and vice versa.
To see this relation in the present approach, we have computed the volume integral of the potential
\bea
	W \equiv 4 \pi \int r^2 dr \tilde{U} \left( r \right)
\eea
and take the ratios of $W$'s with different $F_{\pi}$'s.
In our parameter sets 1 and 2 with  $F_{\pi} = 129 \ \mathrm{MeV}$ and 186 MeV, we find the ratio
\bea
	\displaystyle{\frac{W \left( F_{\pi} = 129 \ \mathrm{MeV} \right)}{W \left( F_{\pi} = 186 \ \mathrm{MeV} \right)}} \sim \displaystyle{\frac{5}{2}} \sim 2.5
\eea
which is compatible with $\left( 186/ 129 \right)^2 \sim 15/7 \sim 2.1$.
A small difference is considered due to the violation of SU(3) in the present ansatz of the bound state approach.

\subsection{Comparisons with the Callan-Klebanov approach}
\label{sec:Comparisons with the CK}
In this subsection, we compare our results with those of Callan-Klevanov (CK).
In thier approach for the hedgehog kaon system, the lowest bound state appears in the p-wave rather 
than s-wave with a strong binding energy of order hundred MeV~\cite{Callan-Klebanov 1, Callan-Klebanov 2, Itzhaki}.
We show the some results for parameter set 3 in Table~\ref{tab:comparison with CK approach}, where $l$ is the kaon orbital angular momentum and $l_{eff}$ is the effective angular momentum defined by~\cite{Callan-Klebanov 1, Callan-Klebanov 2}
\bea
	l_{eff} \left( l_{eff} + 1 \right) = l \left( l + 1 \right) + 4 \bm{I} \cdot \bm{L} + 2.
\label{eq:effective angular momentum}
\eea
The p-wave bound state corresponds to $\Lambda \left(1116 \right)$ and the s-wave to $\Lambda \left(1405 \right)$ in the CK approach~\cite{Callan-Klebanov 1, Callan-Klebanov 2}.
From Table~\ref{tab:comparison with CK approach}, we find that the kaon radii, $\left< r^2_K \right>^{1/2}$, are about 0.5 fm for the p-wave and 0.9 fm for the s-wave in the CK approach~\footnote{We have performed numerical calculations in the CK approach.}, which are substantially smaller than those of our present approach.  
These results for small radii seem consistent with their interpretation of the kaon hedgehog system as the strange quark and diquark system for hyperons.
\begin{table}[htb]
	\begin{center}
		\caption{Comparisons between the CK and our approaches}
			\begin{tabular}{| c | c | c | c || c | c | c || c |} \hline
				\multicolumn{4}{|c||}{Callan-Klebanov approach} 			 & \multicolumn{3}{|c||}{Our approach}			     & Physical state \\ \hline
	  			$l$ & $l_{eff}$ & B. E. [MeV] & $\left< r^2_{K} \right>^{1/2}$ [fm] & $l$ & B. E. [MeV] & $\left< r^2_{K} \right>^{1/2}$ [fm] &  \\ \hline
				0    & 1 	      & 61.7 	    & 0.93 					     	 & 0	  & 32.9	        & 1.18			      		     & $\Lambda \left( 1405 \right)$ \\ \hline 
				1    & 0           & 326.6 	    & 0.54 						 & ---  & --- 	        & ---	  		 	 	 	     & $\Lambda \left( 1116 \right)$ \\ \hline
			\end{tabular}	
\label{tab:comparison with CK approach}
	\end{center}
\end{table}

\begin{figure}[htb]
	\begin{center}
		\includegraphics[width=100mm]{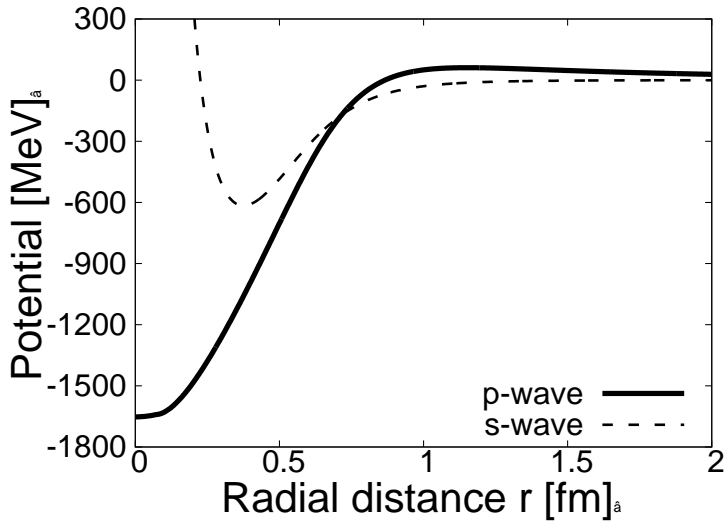}
	\end{center}
 	\caption{The $\bar{K}N$ potentials obtained from the CK approach.}
\label{fig:CK potentials}
\end{figure}

In Figure~\ref{fig:CK potentials}, we show the potentials for the s- and p-wave bound states in the CK approach, which are defined similarly to the one of Eq.~(\ref{eq:potential value}).
For p-wave, the potential has a strong attraction at the origin.
This causes the strong bound state as the ground state in the p-wave.
For s-wave, we see a repulsive component toward the origin.
This is caused by the centrifugal-like component due to the effective angular momentum $l_{eff}$.
A similar structure is seen in our potential in Figure~\ref{fig:potential}.
We consider that the presence of the centrifugal-like potential in the CK approach is related to the presence of the repulsive core in our approach.

\section{Summary}
In this paper, we have constructed a new method for the study of kaon-nucleon systems and their interactions in the Skyrme model based on the bound state approach which Callan and Klebanov proposed~\cite{Callan-Klebanov 1, Callan-Klebanov 2}.
In our approach, we first quantize the hedgehog ansatz to generate the physical nucleon and introduce kaon fluctuations around it.
This is the different point from the Callan-Klebanov approach, where they first introduce the kaon fluctuations around the hedgehog, and then the kaon-hedgehog system is quantized for hyperons.
Although our method does not obey $1/N_c$ expansion systematically, we consider it suitable for a kaon bound system of small binding energy of order ten MeV or less.

As a general structure of interaction between isoscalar-pseudoscalar kaon and isospinor-spinor nucleon, the obtained potential contains central and spin-orbit terms with and without isospin dependence.
A nontrivial finding is that there is either repulsion or attraction proportional to $1/r^2$ for small $r$, depending on the kaon partial wave. 
For $l = 0$, the resulting potential turns out to contain the short range repulsion and the middle range attraction.
Consequently, the kaon bound states obtain a weak binding energy.
The presence of the repulsion should have an influence for the properties of high density kaonic nuclear matter.  
When $1/r^2$ term gives an attraction $\left( I = 0, l = 1, 2, 3, 4, J_{total} = l - 1/2 \right)$, the system becomes unstable.
The present method, however, should not be applied to such a situation, where we need more microscopic approach.  

In the present paper, we have focused our discussion on possible bound states.
An extension to continuum states for kaon nucleon scattering is rather straightforward.
As for kaon scatterings, Karliner and Mattis made a detailed investigation especially for higher partial waves~\cite{Karliner 1, Karliner 2}.
We plan to study scattering states from lower to higher partial waves.
So far we have considered only kaon-nucleon channel, but it is well known that $\pi \Sigma$ channel is also important especially for the discussion of $\Lambda \left( 1405 \right)$.
It is another interesting extension of the present study, which we hope to report elsewhere.

\begin{acknowledgments}
We thank Noriyoshi Ishii and Takuya Sugiura for useful discussions.
This work is supported in part by the Grant-in-Aid  for Science Research (C) 26400273.
\end{acknowledgments}

\appendix
\section{}
In this appendix, we show an outline to derive the potential Eq.~(\ref{eq:potential}).
Substituting our ansatz Eq.~(\ref{ans:our ansatz}) for the lagrangian Eq.~(\ref{eq:lagrangian}), and expanding it up to second order of the kaon fields, we obtain the following lagrangian
\bea
	L = L_{SU(2)} + L_{KN},
\eea
where
\bea
	L_{SU(2)} &=& \displaystyle{\frac{1}{16}} {F_{\pi}}^2 \mathrm{tr} \left[ \del_{\mu} \tilde{U}^{\dag} \del^{\mu} \tilde{U} \right] 
				+\displaystyle{\frac{1}{32e^2}} \mathrm{tr} \left[ \del_{\mu} \tilde{U} \tilde{U}^{\dag}, \del_{\nu} \tilde{U} \tilde{U}^{\dag} \right]^2,
\label{eq:SU(2) Skyrme lagrangian}
\eea
and
\bea
	L_{KN}
		&=& \left( D_{\mu}K\right)^{\dag}D^{\mu}K 
				- K^{\dag} a_{\mu}^{\dag} a^{\mu} K 
				- m_K^2 K^{\dag} K \nonum \\
		&&+ \displaystyle{\frac{1}{(e F_{\pi})^2}} \left\{ -  K^{\dag}K\mathrm{tr}\left[ \del_{\mu} \tilde{U} \tilde{U}^{\dag}, \del_{\nu} \tilde{U} \tilde{U}^{\dag}\right]^2 
			       - 2\left( D_{\mu}K\right)^{\dag}D_{\nu}K \mathrm{tr}\left( a^{\mu}a^{\nu}\right) \right.\nonumber \\
			  && \hspace{2cm} \left. - \displaystyle{\frac{1}{2}}\left( D_{\mu}K\right)^{\dag}D^{\mu}K \mathrm{tr}\left( \del_{\nu} \tilde{U}^{\dag} \del^{\nu} \tilde{U} \right) 
			       + 6\left( D_{\nu}K\right)^{\dag} \left[ a^{\nu},a^{\mu}\right]D_{\mu}K \right\} \nonum \\
		 && +  \displaystyle{\frac{3 i}{F_{\pi}^2}} B^{\mu} \left[ \left( D_{\mu} K \right)^{\dag} K - K^{\dag}  \left( D_{\mu} K \right) \right].
\label{eq:KN lagrangian}
\eea
In these equations, we have defined 
\bea
	\tilde{U} = A \left( t \right) U_{H} A^{\dag} \left( t \right), \ \ \ \ \tilde{\xi} = A \left( t \right) \xi A^{\dag} \left( t \right),
\eea
\bea
	D_{\mu} K = \del_{\mu} K + v_{\mu} K,
\eea
\bea
	v_{\mu} &=& \displaystyle{\frac{1}{2}} \left( \tilde{\xi}^{\dag} \del_{\mu} \tilde{\xi} + \tilde{\xi} \del_{\mu} \tilde{\xi}^{\dag} \right), \\
	a_{\mu} &=& \displaystyle{\frac{1}{2}} \left( \tilde{\xi}^{\dag} \del_{\mu} \tilde{\xi} - \tilde{\xi} \del_{\mu} \tilde{\xi}^{\dag} \right),
\eea
where the hedgehog ansatz $U_H$ and $\xi$ are given by Eq.~(\ref{eq:def of Hg configuration}), and $B^{\mu}$ is the baryon current which is given by~\cite{ANW}
\bea
	B^{\mu} = - \displaystyle{\frac{\varepsilon^{\mu \nu \alpha \beta}}{24 \pi^2}}  
				\mathrm{tr} \left[ \left( U_H^{\dag} \del_{\nu} U_H \right) \left( U_H^{\dag} \del_{\alpha} U_H \right) \left( U_H^{\dag} \del_{\beta} U_H \right) \right].
\eea

From Eq.~(\ref{eq:KN lagrangian}), we derive the equation of motion for the kaon Eq.~(\ref{eq:equation of motion for kaon}) and the potential Eq.~(\ref{eq:potential}) with each term given by
\bea
	V_0^c \left( r \right)
		&=& - \displaystyle{\frac{1}{4}} \left( 2 \frac{\sin^2 F}{r^2} + (F')^2 \right) 
			+ 2 \displaystyle{\frac{s^4}{r^2}}
			+ \left[ 1 + \displaystyle{\frac{1}{\left( e F_{\pi} \right)^2}} \left( F'^2 + \displaystyle{\frac{\sin^2 F}{r^2}} \right)  \right] \displaystyle{\frac{l \left( l + 1 \right)}{r^2}}  \nonum \\
		&& - \displaystyle{\frac{1}{\left( e F_{\pi} \right)^2}}  \left[ 2 \displaystyle{\frac{\sin^2 F}{r^2}} \left( \displaystyle{\frac{\sin^2 F}{r^2}} - 2 (F')^2 \right) 
												   - 2 \displaystyle{\frac{s^4}{r^2}} \left( F'^2 + \displaystyle{\frac{\sin^2 F}{r^2}} \right) \right] \nonum \\
		&& + \displaystyle{\frac{1}{\left( e F_{\pi} \right)^2}} \displaystyle{\frac{6}{r^2}} \left[ \displaystyle{\frac{s^4 \sin^2 F}{r^2}} - \displaystyle{\frac{d}{dr}} \left\{ s^2 \sin F F' \right\} \right] \nonum \\
		&& + \displaystyle{\frac{2 E}{\Lambda}} s^2 \left[ 1 + \displaystyle{\frac{1}{\left( e F_{\pi} \right)^2}} \left( F'^2 + \displaystyle{\frac{5}{r^2}} \sin^2 F \right) \right] \nonum \\
		&& + \displaystyle{\frac{3}{\left( e F_{\pi} \right)^2}} \displaystyle{\frac{1}{r^2}} \displaystyle{\frac{d}{dr}} \left[ r^2 \left( \displaystyle{\frac{E F' \sin F }{\Lambda}}  \right)  \right] 
			\pm \displaystyle{\frac{3}{\pi^2 F_{\pi}^2}} \displaystyle{\frac{\sin^2 F}{r^2}} F'  \left( E - \displaystyle{\frac{s^2}{\Lambda}} \right),
\label{eq:V0c}
\eea
\bea
	V_{\tau}^c \left( r \right)
		&=& \displaystyle{\frac{8 E}{3 \Lambda}} s^2 \left[ 1 + \displaystyle{\frac{1}{\left( e F_{\pi} \right)^2}}  \left( F'^2 + \displaystyle{\frac{4}{r^2}} \sin^2 F \right) \right]
			+ \displaystyle{\frac{4}{\left( e F_{\pi} \right)^2}} \displaystyle{\frac{1}{r^2}} \displaystyle{\frac{d}{dr}} \left[ r^2 \left( \displaystyle{\frac{ E F' \sin F}{\Lambda}} \right)  \right], \nonum \\
\label{eq:Vtauc}
\eea
\bea
	V_0^{LS} \left( r \right)
		&=& \displaystyle{\frac{1}{\left( e F_{\pi} \right)^2}} \displaystyle{\frac{2E \sin^2 F}{\Lambda r^2}} 
			\pm \displaystyle{\frac{3}{F_{\pi}^2 \pi^2}} \displaystyle{\frac{\sin^2 F}{\Lambda r^2}} F',
\label{eq:V0LS}
\eea
and
\bea
	V_{\tau}^{LS} \left( r \right)
		&=& - \left[ 1 + \displaystyle{\frac{1}{\left( e F_{\pi} \right)^2}}  \left( F'^2 + 4 \displaystyle{\frac{\sin^2 F}{r^2}} \right) \right] \displaystyle{\frac{16 s^2}{3 r^2}} 
			- \displaystyle{\frac{1}{\left( e F_{\pi} \right)^2}}\displaystyle{\frac{8}{r^2}} \left[ \displaystyle{\frac{d}{dr}}\left( \sin F F'\right) \right],
\label{eq:VtauLS}
\eea
where
\bea
	s = \sin \left( F \left( r \right) / 2 \right), 
\eea
and
\bea
	F' = dF \left( r \right)/dr.
\eea
The moment of inertia $\Lambda$ is given by
\bea
	\Lambda = \displaystyle{\frac{2 \pi}{3}} F_{\pi}^2 \int dr r^2 \sin^2 F \left[ 1 + \displaystyle{\frac{4}{\left( e F_{\pi} \right)^2}} \left( F'^2 + \displaystyle{\frac{\sin^2 F}{r^2}} \right) \right].
\label{eq:moment of inertia}
\eea
The last terms of Eq.~(\ref{eq:V0c}) and Eq.~(\ref{eq:V0LS}) are derived from the Wess-Zumino term, which is attractive for the $\bar{K}N$ potential and repulsive for the $KN$ potential.
These equations are general for any partial waves of the kaon.
For instance, the $s$-wave potential is obtained by setting $l = 0$ and removing the terms including $J_{KN}$ in Eq.~(\ref{eq:potential}).

Now we discuss two features of the potential, the relation with the Weinberg-Tomozawa interaction (WT interaction)~\cite{WT 1, WT 2} and the short range behaviors.
To see the essential aspect of the WT interaction for the kaon-nucleon interaction, let us look at the leading contribution from the pion fields, derived from the kinetic term where the covariant derivative in the lagrangian Eq.~(\ref{eq:KN lagrangian}) is given by
\bea
	D_{\mu} K = \del_{\mu} K + v_{\mu} K,
\label{eq:def of covariant derivative with pionic soliton}
\eea
with
\bea
	v_{\mu} &=& \displaystyle{\frac{1}{2}} \left( \xi^{\dag} \del_{\mu} \xi + \xi \del_{\mu} \xi^{\dag} \right),
	\ \ \ \ U_{\pi} = \xi^2,
\eea
and $U_{\pi}$ given by Eq.~(\ref{eq:def of pionic soliton}).
Picking up the terms of order $\mathcal{O} \left( \pi^2 \right)$, we find
\bea
	L_{WT} &=& \displaystyle{\frac{2 i}{F_{\pi}^2}} \left[ \del_{\mu} K^{\dag} \displaystyle{\frac{\bm{\tau}}{2}} K  - K^{\dag} \displaystyle{\frac{\bm{\tau}}{2}} \del_{\mu} K \right]
				 \cdot \left( \bm{\pi} \times  \del^{\mu} \bm{\pi} \right) \nonum \\
			&=& \displaystyle{\frac{1}{2 F_{\pi}^2}} \left( \del_{\mu} K^{\dag} \left[ \pi, \del^{\mu} \pi  \right] K
										- K^{\dag} \left[ \pi, \del^{\mu} \pi  \right] \del_{\mu} K \right),
\label{eq:WT type lagrangian}
\eea
where we have defined $\pi = \bm{\tau} \cdot \bm{\pi}$.
We note that the first line of Eq.~(\ref{eq:WT type lagrangian}) takes the form of the product of the isospin vector currents of the kaon and the pion fields.

For the kaon and nucleon interaction, we first substitute the hedgehog ansatz for the pion field,
\bea
	\pi = \bm{\tau} \cdot \bm{\pi} = \displaystyle{\frac{F_{\pi}}{2}} F\left( r \right) \bm{\tau} \cdot \hat{r}.
\label{eq:relation between pionic soliton and hedgehog ansatz}
\eea
Then, we rotate the hedgehog ansatz in SU(2) space
\bea
	F\left( r \right) \bm{\tau} \cdot \hat{r} \rightarrow F\left( r \right) A \left( t \right) \bm{\tau} \cdot \hat{r} A^{\dag} \left( t \right), \ \ \ A \left( t \right) \in \mathrm{SU(2)}.
\label{eq:rotating isospin matrix}
\eea
Substituting Eq.~(\ref{eq:relation between pionic soliton and hedgehog ansatz}) and Eq.~(\ref{eq:rotating isospin matrix}) for Eq.~(\ref{eq:WT type lagrangian}), the leading contribution of the WT interaction is
\bea
	L_{WT} \simeq \displaystyle{\frac{i}{12 \Lambda}} F^2 \left( r \right)  
					\left[ \del_{0} K^{\dag} \left( \bm{\tau}^K \cdot \bm{\tau}^N \right) K - K^{\dag} \left( \bm{\tau}^K \cdot \bm{\tau}^N \right) \del_{0} K \right],
\label{eq:lo WT type lagrangian}
\eea
where $\Lambda$ is given by Eq.~(\ref{eq:moment of inertia}).

On the other hand, in our approach, we obtain the following contributions from the kinetic term in the lagrangian Eq.~(\ref{eq:KN lagrangian}) using our ansatz 
\bea
	&&\displaystyle{\frac{i}{3 \Lambda}} \sin^2 \left( \displaystyle{\frac{F\left( r \right)}{2}} \right) 
		\left[ \del_{0} K^{\dag} \left( \bm{\tau}^K \cdot \bm{\tau}^N \right) K - K^{\dag} \left( \bm{\tau}^K \cdot \bm{\tau}^N \right) \del_{0} K \right] \nonum \\
		&\simeq&\displaystyle{\frac{i}{12 \Lambda}} F^2 \left( r \right)  
			\left[ \del_{0} K^{\dag} \left( \bm{\tau}^K \cdot \bm{\tau}^N \right) K - K^{\dag} \left( \bm{\tau}^K \cdot \bm{\tau}^N \right) \del_{0} K \right].
\label{eq:WT type lagrangian from our approach}
\eea
Comparing Eq.~(\ref{eq:lo WT type lagrangian}) with Eq.~(\ref{eq:WT type lagrangian from our approach}), we find that they coinside each other up to the leading order of $F^2 \left( r \right)$.
 
Next, we consider how the potential behaves near the origin.
From the equation of motion for $F \left( r \right)$~\cite{ANW}, the behavior of $F \left( r \right)$ near the origin is given by
\bea
	F \left( r \simeq 0 \right) = \pi - ar,
\label{eq:profile function near the origin}
\eea 
\begin{table}[htb]
	\begin{center}
		\caption{Behaviors of the potential near the period}
			\begin{tabular}{| c | c | c |} \hline
	  					    & $J_{tot} = l - 1/2$		                    	  & $J_{tot} = l + 1/2$  \\ \hline
				$I_{tot} = 0$ & attractive $\left( l  = 1, 2, 3, 4 \right)$      	  & repulsive \\ 
						    & repulsive ($ l  = 0,$ and $, 5, 6, 7, \cdots$ ) & \\ \hline
				$I_{tot} = 1$ & repulsive               				    	  & repulsive \\ \hline
			\end{tabular}
			\label{tab:behavior of potential}
	\end{center}
\end{table}
where $a$ is a constant which is determined by the soliton profile function $F \left( r \right)$.
Using Eq.~(\ref{eq:profile function near the origin}), the potential reduces to 
\bea
	V \left( r \simeq 0 \right)
		&=& \displaystyle{\frac{2}{r^2}} 
			+ \displaystyle{\frac{a^2}{(e F_{\pi})^2}} \displaystyle{\frac{4}{r^2}}
			+ \left[ 1 + \displaystyle{\frac{2 a^2}{(e F_{\pi})^2}} \right] \displaystyle{\frac{l \left( l + 1 \right)}{r^2}} \nonum \\
			&& - \left[ 1 + \displaystyle{\frac{5 a^2}{(e F_{\pi})^2}} \right] \displaystyle{\frac{16}{3 r^2}} J_{KN} I_{KN}
			- \displaystyle{\frac{a^2}{(e F_{\pi})^2}} \displaystyle{\frac{8}{r^2}}  J_{KN} I_{KN} .
\label{eq:potential for r=0}
\eea
Whether this potential becomes either attractive or repulsive depends on the total isospin $I_{tot}$ and the total spin $J_{tot}$, as shown in Table~\ref{tab:behavior of potential}.



\begin{thebibliography}{99}
\bibitem{KbarN bound state 1}
R.~H.~Dalitz and S.~F.~Tuan,
Phys.\ Rev.\ Lett.\  {\bf 2}, 425 (1959).
doi:10.1103/PhysRevLett.2.425

\bibitem{KbarN bound state 2}
R.~H.~Dalitz and S.~F.~Tuan,
Annals Phys.\  {\bf 10}, 307 (1960).
doi:10.1016/0003-4916(60)90001-4
  
\bibitem{few-body system 1}
A.~N.~Ivanov, P.~Kienle, J.~Marton and E.~Widmann,
nucl-th/0512037.

\bibitem{few-body system 2}
N.~V.~Shevchenko, A.~Gal, J.~Mares and J.~Revai,
Phys.\ Rev.\ C {\bf 76}, 044004 (2007)
doi:10.1103/PhysRevC.76.044004
[arXiv:0706.4393 [nucl-th]].

\bibitem{few-body system 3}
Y.~Ikeda and T.~Sato,
Phys.\ Rev.\ C {\bf 76}, 035203 (2007)
doi:10.1103/PhysRevC.76.035203
[arXiv:0704.1978 [nucl-th]].
 
\bibitem{few-body system 4}
A.~Dote, T.~Hyodo and W.~Weise,
Nucl.\ Phys.\ A {\bf 804}, 197 (2008)
doi:10.1016/j.nuclphysa.2008.02.001
[arXiv:0802.0238 [nucl-th]].

\bibitem{few-body system 5}
T.~Nishikawa and Y.~Kondo,
Phys.\ Rev.\ C {\bf 77}, 055202 (2008)
doi:10.1103/PhysRevC.77.055202
[arXiv:0710.0948 [hep-ph]].

\bibitem{yamazaki 1} 
T.~Yamazaki and Y.~Akaishi,
Phys.\ Lett.\ B {\bf 535}, 70 (2002).
doi:10.1016/S0370-2693(02)01738-0

\bibitem{yamazaki 2}
Y.~Akaishi and T.~Yamazaki,
Phys.\ Rev.\ C {\bf 65}, 044005 (2002).
doi:10.1103/PhysRevC.65.044005

\bibitem{chiral potential 1}
B.~Borasoy, R.~Nissler and W.~Weise,
Eur.\ Phys.\ J.\ A {\bf 25}, 79 (2005)
doi:10.1140/epja/i2005-10079-1
[hep-ph/0505239].

\bibitem{chiral potential 2}
T.~Hyodo and W.~Weise,
Phys.\ Rev.\ C {\bf 77}, 035204 (2008)
doi:10.1103/PhysRevC.77.035204
[arXiv:0712.1613 [nucl-th]].

\bibitem{chiral potential 3}
Y.~Ikeda, T.~Hyodo and W.~Weise,
Nucl.\ Phys.\ A {\bf 881}, 98 (2012)
doi:10.1016/j.nuclphysa.2012.01.029
[arXiv:1201.6549 [nucl-th]].

\bibitem{WT 1}
S.~Weinberg,
Phys.\ Rev.\ Lett.\  {\bf 17}, 616 (1966).
doi:10.1103/PhysRevLett.17.616

\bibitem{WT 2}
Y.~Tomozawa,
Nuovo Cim.\ A {\bf 46}, 707 (1966).
doi:10.1007/BF02857517

\bibitem{Skyrme model 1}
T.~H.~R.~Skyrme,
Proc.\ Roy.\ Soc.\ Lond.\ A {\bf 260}, 127 (1961).
doi:10.1098/rspa.1961.0018

\bibitem{Skyrme model 2}
J.~K.~Perring and T.~H.~R.~Skyrme,
Nucl.\ Phys.\  {\bf 31}, 550 (1962).
doi:10.1016/0029-5582(62)90774-5

\bibitem{Skyrme model 3}
T.~H.~R.~Skyrme,
Nucl.\ Phys.\  {\bf 31}, 556 (1962).
doi:10.1016/0029-5582(62)90775-7

\bibitem{Callan-Klebanov 1}
C.~G.~Callan, Jr. and I.~R.~Klebanov,
Nucl.\ Phys.\ B {\bf 262}, 365 (1985).
doi:10.1016/0550-3213(85)90292-5
 
 \bibitem{Callan-Klebanov 2}
C.~G.~Callan, Jr., K.~Hornbostel and I.~R.~Klebanov,
Phys.\ Lett.\ B {\bf 202}, 269 (1988).
doi:10.1016/0370-2693(88)90022-6
  
\bibitem{WZ term 1}
J.~Wess and B.~Zumino,
Phys.\ Lett.\ B {\bf 37}, 95 (1971).
doi:10.1016/0370-2693(71)90582-X

\bibitem{WZ term 2}
E.~Witten,
Nucl.\ Phys.\ B {\bf 223}, 422 (1983).
doi:10.1016/0550-3213(83)90063-9

\bibitem{WZ term 3}
E.~Witten,
Nucl.\ Phys.\ B {\bf 223}, 433 (1983).
doi:10.1016/0550-3213(83)90064-0

\bibitem{Ring-Schuck}
P.~Ring and P.~Schuck, {\it The Nuclear Many-Body Problem}
(Springer,\ New York,\ 1980).

\bibitem{SU(3) Skyrme model 1}
M.~Praszalowicz,
Phys.\ Lett.\ B {\bf 158}, 264 (1985).
doi:10.1016/0370-2693(85)90968-2

\bibitem{SU(3) Skyrme model 2}
H.~Yabu and K.~Ando,
Nucl.\ Phys.\ B {\bf 301}, 601 (1988).
doi:10.1016/0550-3213(88)90279-9

\bibitem{zahed}
I.~Zahed and G.~E.~Brown,
Phys.\ Rept.\  {\bf 142}, 1 (1986).
doi:10.1016/0370-1573(86)90142-0
  
\bibitem{ANW}
G.~S.~Adkins, C.~R.~Nappi and E.~Witten,
Nucl.\ Phys.\ B {\bf 228}, 552 (1983).
doi:10.1016/0550-3213(83)90559-X

\bibitem{PDG}
K.~A.~Olive {\it et al.} [Particle Data Group Collaboration],
Chin.\ Phys.\ C {\bf 38}, 090001 (2014).
doi:10.1088/1674-1137/38/9/090001

\bibitem{Itzhaki} 
N.~Itzhaki, I.~R.~Klebanov, P.~Ouyang and L.~Rastelli,
Nucl.\ Phys.\ B {\bf 684}, 264 (2004)
doi:10.1016/j.nuclphysb.2004.02.004
[hep-ph/0309305].
  
\bibitem{Karliner 1} 
M.~Karliner and M.~P.~Mattis,
Phys.\ Rev.\ D {\bf 34}, 1991 (1986).
doi:10.1103/PhysRevD.34.1991

\bibitem{Karliner 2} 
M.~Karliner,
Phys.\ Rev.\ Lett.\  {\bf 57}, 523 (1986).
doi:10.1103/PhysRevLett.57.523


\end{thebibliography}
\end{document}